# Observation of Topological Armchair Edge States
# in Photonic Biphenylene Network

Qi Zhong[1†], Yongsheng Liang[1†], Shiqi Xia[1*], Daohong Song[1, 2], Zhigang Chen[1, 2*]

*[1]The MOE Key Laboratory of Weak-Light Nonlinear Photonics, TEDA Applied Physics Institute and School of Physics, Nankai University, Tianjin 300457, China*

*[2]Collaborative Innovation Center of Extreme Optics, Shanxi University, Taiyuan, Shanxi 030006, China*

*[†]These authors contributed equally to this work*

*[*]Corresponding author: shiqixia@nankai.edu.cn, zgchen@nankai.edu.cn*

## Abstract:

Edge states in 2D materials are vital for advancements in spintronics, quantum computing, and logic transistors. For graphene nanoribbons, it is well known that the zigzag edges can host edge states, but realization of armchair edge states has been challenging without breaking the time-reversal symmetry. Here, by using a photonic analog of recently synthesized graphene-like biphenylene network (BPN), we demonstrate topological in-gap edge states particularly at the armchair edges. Interestingly, several bulk states preserve the characteristics of edge states along the armchair boundaries, manifesting an unusual hybridization between the edge and bulk states. Experimentally, we observe both zigzag and armchair edge states in photonic BPN lattices written in a nonlinear crystal. Furthermore, we clarify the different features of the armchair boundary between the BPN and graphene lattices. Our results demonstrated here may be applicable to carbon-based BPNs and other artificial platforms beyond photonics, holding promise for expanding the application scope of 2D materials.

# 1、Introduction

Graphene and related 2D materials are at the frontier of material science research due to their unique properties and potential applications, particularly in spintronics, quantum computing, and next-generation electronic and photonic devices [1-6]. One of the key areas of interest is the study of edge-dependent states, which have been extensively explored in the context of graphene nanoribbons [7-9]. Edge states are integral to the development of graphene-based devices, as they can affect the electronic properties of the materials. Graphene exhibits four fundamental edge types: zigzag, bearded, and armchair edges [10], along with the twig edge recently reported in synthetic photonic lattices [11]. For carbon-based graphene nanoribbons, only the boundaries with zigzag, armchair or their combinations can be synthesized due to the mechanical instability of atomic dangling bonds [12, 13]. For these two boundaries, the zigzag edge states have been observed, which endow graphene nanoribbons with unique electronic, spintronic, and magnetic properties [7, 14, 15], whereas the armchair edge states are not supported under the tight-binding model. Exploration of 2D materials with armchair and zigzag edge states can leverage the advantages of each edge type to enhance electronic and spintronic properties, as well as adjustable conductivity, which are important for multifunctional nanoscale devices [15-19] and potential applications in next-generation topological and quantum devices [20, 21].

Biphenylene network (BPN), experimentally synthesized as a carbon allotrope with a graphene-like structure [22], has emerged as a new material platform, theoretically explored for magnetic ordering, Lifshitz transition [23], and the formation of topological armchair edge states [24]. While BPN offers a promising structure for realizing armchair edge states, the experimental observation of these edge states has not been achieved. In recent years, synthetic 2D structures, such as photonic crystals and acoustic metamaterials, have been developed to emulate edge states in 2D materials, which offer a much greater control of physical parameters as compared with carbon-based real materials. Among different synthetic structures, photonics lattices have proven invaluable for exploring topological physics, leading to advancements in photonic Floquet topological insulators [25], valley Hall topological insulators [26], and higher-order topological insulators [27-30].

In this work, we theoretically analyze and experimentally demonstrate the existence of topological armchair edge states in BPN nanoribbons using photonic lattices. The topological properties of these edge states are revealed by Zak phase calculations. Moreover, we analyze the underlying physical mechanism for the formation of armchair edge states in BPN, setting them apart from those previously known edge states in graphene. By employing a continuous-wave (cw) laser-writing technique, the desired photonic BPN lattices are constructed, and as a result, we observe edge states on the zigzag and armchair boundaries. This work represents the first direct observation of armchair edge states in any BPN lattice.

## 2、 Theoretical analysis

In two-dimensional lattices, different truncations create distinct boundaries, which in turn can give rise to edge states that reflect the bulk topology. For instance, the graphene lattice can support zigzag, bearded, and twig edge states [11]. In a graphene lattice, there are two sublattices within one unit cell (Fig. 1a1), and it features chiral symmetry ($H\Gamma = -\Gamma H$), where $\Gamma$ is the chiral symmetry operator and $H$ is the Hamiltonian. The Dirac points are located at the zero-energy level (Fig. 1a2). The edge states, protected by the nontrivial winding arising from the Dirac points, are pinned at zero energy ($E = 0$), due to the chiral symmetry of graphene. In this case, $\Gamma|\varphi\rangle = \pm|\varphi\rangle$, where $|\varphi\rangle$ is the eigenvector with zero eigenvalues, and the edge states in real space are distributed only on one type of sublattice. However, this criterion conflicts with the energy distribution required for stable transmission under the armchair edge condition in graphene. The reasoning behind this is twofold. Firstly, the armchair edge condition exhibits trivial winding in momentum space. Secondly, suppose an edge state $|\varphi_{\text{edge}}\rangle$ with zero-energy eigenvalue can appear at the graphene armchair edge, it cannot distribute stably only on one type of sublattice due to the spatial geometry of the lattice, as illustrated in Fig. 1a1. In other words, $H|\varphi_{\text{edge}}\rangle \neq E|\varphi_{\text{edge}}\rangle$, where $E = 0$. Interestingly, the BPN lattice has the same armchair boundary as the graphene lattice but it can support topological edge states along the armchair boundary (Fig. 1b1), which reflects different spatial geometry and topological properties of the two lattices.

The BPN is a six-membered ring network formed by the translation of the six-site unit cells as shown in Fig. 1b1. Under the tight-binding approximation, the Hamiltonian can be written as [24]:

$$H(\mathbf{k}) = \begin{pmatrix} 0 & t & 0 & 0 & te^{i\mathbf{k}\cdot\mathbf{a_2}} & t \\ t & 0 & t & te^{i\mathbf{k}\cdot\mathbf{a_2}} & 0 & 0 \\ 0 & t & 0 & t & 0 & te^{i\mathbf{k}\cdot\mathbf{a_1}} \\ 0 & te^{-i\mathbf{k}\cdot\mathbf{a_2}} & t & 0 & t & 0 \\ te^{-i\mathbf{k}\cdot\mathbf{a_2}} & 0 & 0 & t & 0 & t \\ t & 0 & te^{-i\mathbf{k}\cdot\mathbf{a_1}} & 0 & t & 0 \end{pmatrix} \quad (1)$$

where $\mathbf{a_1}$ and $\mathbf{a_2}$ are the lattice vectors and $t$ is the nearest neighbor couplings between sites. The topological properties are described by the topological invariant of bulk Hamiltonian and usually manifested as the edge states along boundaries, known as the bulk-boundary correspondence. Notably, Zak phase, as a useful topological invariant, can be applied for the prediction of edge states. The Zak phase of the $n$th sub-band is defined as [30-32]:

$$Z_{k_{x(y)}}^n\big(k_{y(x)}\big) = \int_C dk_{x(y)} A_{k_{x(y)}}^n(k_x, k_y). \quad (2)$$

It is given as the integral of the Berry connection $A_{k_{x(y)}}^n(k_x, k_y) = i\langle u_{n,\mathbf{k}} | \partial_{k_{x(y)}} | u_{n,\mathbf{k}}\rangle$ along a closed path in the Brillouin zone (BZ) [31]. In a discrete form, the Eq. (2) can be written as:

$$Z_{k_{x(y)}}^n\big(k_{y(x)}\big) = \mathrm{Arg}\left(\prod_{j=1}^{N_0} \langle u_{n,k_{x(y)}} | u_{n,k_{x(y)}+\Delta k_{x(y)}}\rangle\right), \quad (3)$$

where $N_0$ represents the number of eigenstates along the integration path $C$ in the $k_x$ or $k_y$ direction. The occurrence of topological edge states in a BPN ribbon can be characterized by the Zak phase summed over all sub-bands lying below the gap, or by the mod $(\sum_{n=1}^m Z_k^n(k), 2\pi)$, where $m$ is the number of bands below the gap.

Different from the graphene structure, the Dirac points in a BPN lattice are not positioned at the zero-energy level ($E \neq 0$) (Fig. 1b2). Therefore, $\Gamma|\varphi\rangle \neq \pm|\varphi\rangle$, where $\Gamma$ is the chiral symmetry operator for the BPN, and $|\varphi\rangle$ is the eigenvector with non-zero eigenvalues. It prevents the edge states from being confined to a single class of sublattices, thus avoiding the constraints typically imposed by chiral symmetry at zero energy. The edge states can be distributed at different types of sublattices (Fig. 1b1), satisfying the corresponding Hamiltonian of the system ($H|\varphi_{\mathrm{edge}}\rangle = E|\varphi_{\mathrm{edge}}\rangle$). Therefore, an armchair boundary in BPN can support stable edge states that align with the spatial geometry of the system.

The armchair BPN ribbon supports two edge states which are fully within the gap and here they are referred to as armchair edge state-I and armchair edge state-II (Fig. 1b1). The corresponding Zak phase $Z_{k_y}^n(k_x)$ is integrated along the path indicated by the red arrow in Fig. 2a1, and the values of the Zak phase are shown in Fig. 2a3, where $n$ denotes the band number. Since the summation of the Zak phase for the first two sub-bands equals $\pi$, it leads to the formation of the topological armchair edge state-I in the gap between the second and third sub-bands (Fig. 2a2). The presence and absence of edge states in other band gaps follow the same rules. The analysis and calculations highlight the relationship between the topological phase and the presence of edge states. It is worth noting that the system exhibits a nontrivial Zak phase below the gap between the first and second (fifth and sixth) sub-bands (Fig. 2a3). However, instead of edge states residing in the yellow-shaded regions within the gap, several bulk states are found to retain the characteristics of edge states. This phenomenon indicates an unusual hybridization between edge and bulk states. Further details can be found in the Supplementary Information (SI) and they merit further exploration in the future.

On the other hand, regarding the BPN ribbon with a zigzag boundary, the integration path $C$ in Eq. (2) is represented by a cyan arrow (Fig. 2b1). Following the same procedure, we can find that there are two chiral edge states, named as zigzag edge state-I and zigzag edge state-II (Fig. 2b2). Notably, both armchair and zigzag boundaries in BPN ribbon support topological edge states. In what follows, we shall discuss experimental results of these distinct types of edge states demonstrated in a photonic BPN lattice.

## 3、Experimental Results

In our experiment, the photonic BPN is constructed using continuous-wave laser writing technique with a wavelength $\lambda = 488\,nm$ [33] (details can be found in SI). The propagation of light in a photonic lattice is governed by the Schrödinger-type paraxial wave equation [34]:

$$i\frac{\partial \Psi(x,y,z)}{\partial z} = -\left(\frac{1}{2k_0}\nabla^2 + \frac{k_0 \Delta n(x,y)}{n_0}\right)\Psi(x,y,z) \qquad (4)$$

where $\Psi(x,y,z)$ is the electric field envelope of a probe beam; $z$ is the propagation distance; $\nabla^2 = \partial_x^2 + \partial_y^2$ is the transverse Laplacian operator; $k_0$, $n_0$ and $\Delta n$ represent the wavenumber, the background refractive index of the nonlinear strontium barium niobate (SBN:61) crystal, and the change in refractive index, respectively.

A photonic biphenylene network with 5×5 unit cells is constructed in the experiment, and is used to demonstrate the topological edge states (Fig. 3a). The photonic lattice has a nearest neighbor waveguide spacing of $35\ \mu m$, with the armchair (zigzag) boundary aligning with the $x(y)$-axis. To match the mode distributions of edge states, a spatial light modulator (SLM) is employed to precisely adjust the intensity and phase of the input beams. These input beams occupy four sites within a unit cell and exhibit an exponential decay from the edge into the bulk (Fig. 3b). For the excitation of armchair edge state-I, the neighboring sites of input beams in one unit cell have a phase difference of $\pi$ (inset in Fig. 3c1). In contrast, the input beams for the armchair edge state-II have the same phase between neighboring sites (inset in Fig. 3d1). All probe beams are launched towards the edge of the 1D BZ, as indicated by the corresponding Fourier spectra (insets in Fig. 3c1, d1). After $20\ mm$ of propagation, the output beams remain localized at the initially excited waveguides, without noticeable coupling to the neighboring sites, demonstrating stable edge state propagation (Fig. 3c1, d1). For comparison, similar probe beams with the same intensity pattern as those used for the armchair edge states, but with a uniform phase distribution, are sent into the lattice (inset in Fig. 3e1). The output shows clear delocalization: light couples to the waveguides marked by the white shadows which initially had no distribution (Fig. 3e1). The corresponding simulation results for the two edge states are shown in Fig. 3c2 and 3d2. These results suggest that both armchair edge states (I and II) remain stable after long-distance propagation. In contrast, the probe beams with a uniform phase distribution radiate throughout the bulk (Fig. 3e2) and cannot be localized.

Under the zigzag edge condition along the y-direction, the zigzag edge state-I occupies five sites within a unit cell (Fig. 4a). The probe beams are again constructed using the SLM in the experiment, and are launched towards $k_y = 0$ in momentum space (see corresponding Fourier spectra in the inset of Fig. 4c1). The probe beams remain intact after $20\ mm$ long propagation (Fig. 4c1), which agrees well with the associated simulation results after longer propagation distance (in Fig. 4c2). The probe beams with a uniform phase distribution are also applied for a direct comparison (Fig. 4d1). The output beams deteriorate in the latter case, and light couples to the neighboring, previously unexcited sites after propagation, as marked by the white shadows (Fig. 4d1). The long-distance simulation results show that the probe beams are severely deformed with leakage into the bulk (Fig. 4d2). The simulation results for zigzag edge state-II can be found in SI (Fig. S5). We have thus demonstrated the presence of topological edge states along both armchair and zigzag boundaries in the photonic BPN for the first time.

# 4、Conclusion

In summary, we have theoretically and experimentally demonstrated the existence of topological edge states along both armchair and zigzag boundaries in photonic BPN lattices, providing key insights into the topological origins of these states. Notably, a subset of the armchair edge states with nontrivial winding numbers can hybridize with bulk states, revealing unusual edge-bulk interactions that merit further exploration. Recent studies indicate that flux-dressed BPN with higher orbital couplings can sustain compact topological edge states [35], broadening the scope of potential new phenomena in these systems. Unlike boundaries with dangling atoms, armchair edge states are more robust and easier to realize in condensed matter systems, enhancing their practical value for applications in electronic materials.

With the successful synthesis of 2D carbon-based BPN structures [22], our findings not only deepen the understanding of edge states in graphene-like and anisotropic materials but also open pathways for a variety of technological applications. These include advanced circuitry, acoustic crystals, and resonator systems [36-39], along with potential implications for spintronics and quantum computing [40, 41]. The stability and tunability of topological edge states in these structures highlight their relevance for developing next-generation functional materials and devices.

**Acknowledgements**

The authors would like to thank Hrovje Buljan for valuable discussions and insights. This work was supported by National Key R&D Program of China (No. 2022YFA1404800); the National Nature Science Foundation of China (No. 12134006, 12274242, and 12474387); the Natural Science Foundation of Tianjin (No. 21JCJQJC00050) and the 111 Project (No. B23045) in China.

**Data availability**

Data supporting key conclusions of this work are included within the article and Supplementary Information. The data that support the findings of this study are available from the corresponding author upon reasonable request

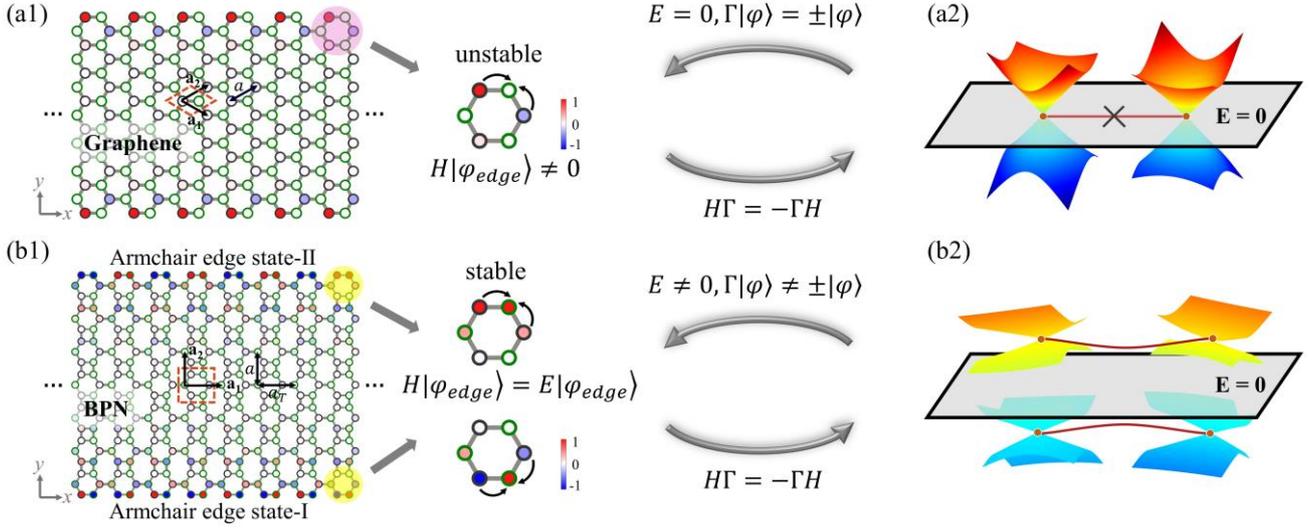

**Figure 1. Illustration of unstable and stable armchair edge states in graphene and biphenylene network (BPN).** (a1) Schematic of the graphene lattice, where the dashed line circles the unit cell of the lattice. $\mathbf{a_1} = \left(\frac{\sqrt{3}a}{2}, -\frac{a}{2}\right)$, $\mathbf{a_2} = \left(\frac{\sqrt{3}a}{2}, \frac{a}{2}\right)$ are the lattice vectors. The different types of sublattice in graphene and BPN are indicated by circles with different colors around sublattice. (a2) Graphene has Dirac points at zero energy, leading to $\Gamma|\varphi\rangle = \pm|\varphi\rangle$. The geometry of the system cannot support stable armchair edge states ($H|\varphi_{edge}\rangle \neq 0$). (b1) Schematic of the BPN lattice, where $\mathbf{a_1} = (a_T, 0)$, $\mathbf{a_2} = (0, a)$, $a_T = \frac{3a}{\sqrt{3}+1}$. (b2) The Dirac points in BPN are positioned at non-zero energy levels ($E \neq 0$), and the edge mode can reside on different sublattice sites ($\Gamma|\varphi\rangle \neq \pm|\varphi\rangle$). The BPN has the armchair edge state-I and armchair edge state-II, which satisfies $H|\varphi_{\text{edge}}\rangle = E|\varphi_{\text{edge}}\rangle$. These figures illustrate why edge states exist in the armchair edge of BPN, but not of graphene lattices.

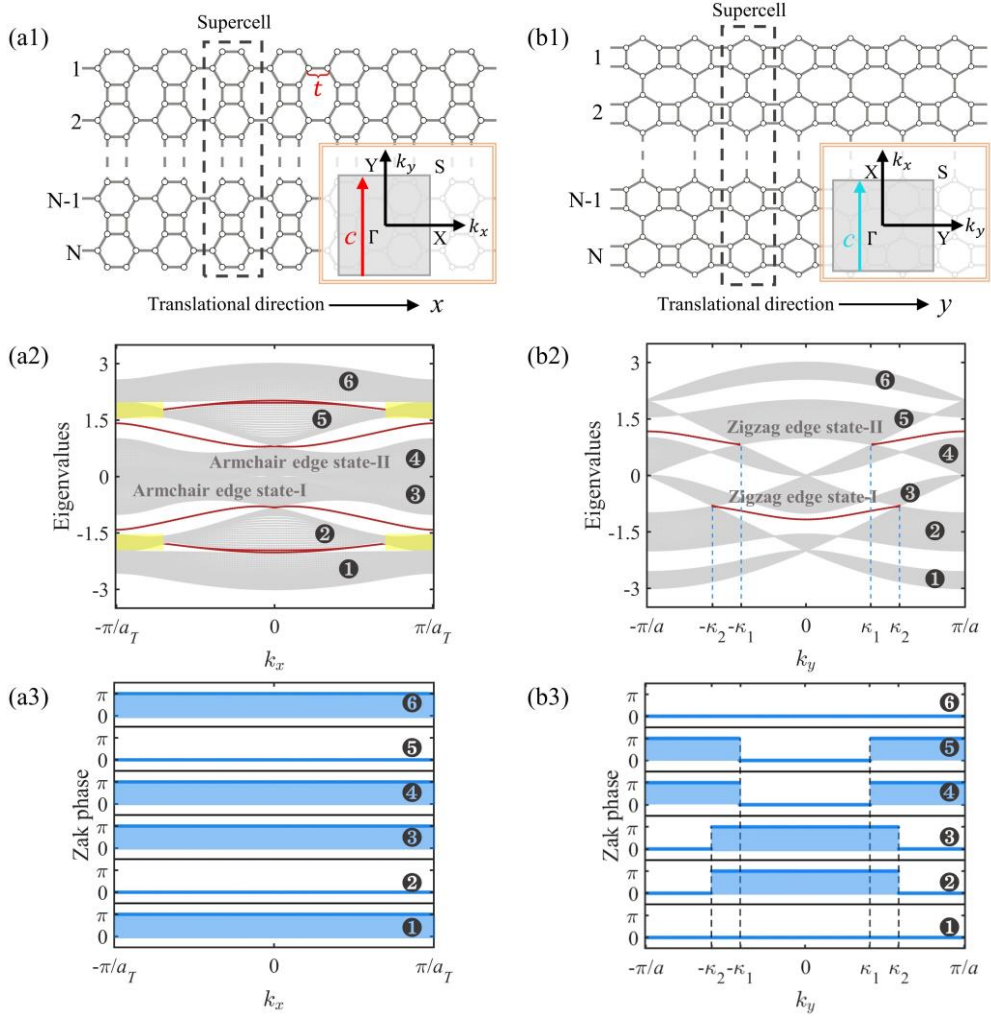

**Figure 2. Topological properties of armchair and zigzag edge states in BPN.** (a1, b1) Schematic lattice structure of the BPN ribbon with armchair boundary (a1) and zigzag boundary (b1), where the grey dashed lines mark the supercell. The nearest neighbor coupling is $t$. The bottom-right inset shows the first Brillouin zone (BZ), where the red and cyan arrows indicate the integration path $C$ for the calculation of the Zak phase. (a2, b2) 1D band structure of the armchair BPN ribbon and zigzag BPN ribbon. Red lines indicate that the topological edge states emerge in the band gaps where the Zak phase is $\pi$. (a3, b3) The corresponding Zak phase of the six sub-bands of the armchair BPN (a3) and zigzag BPN (b3) ribbons. The blue shadows mark the regions where the Zak phase is $\pi$.

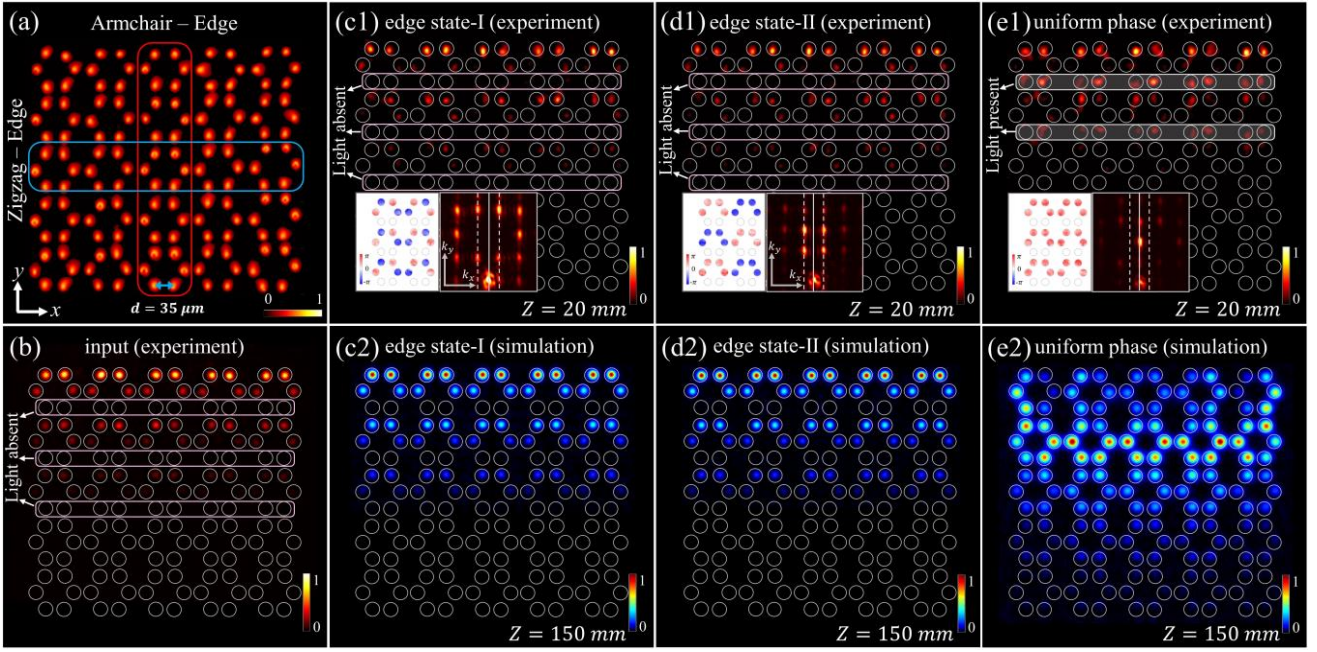

**Figure 3. Experimental demonstration of armchair edge state-I and -II in the photonic BPN.** (a) The photonic BPN lattice established in the experiment, with a distance between two nearest neighbor sites $d = 35\ \mu m$. The armchair boundary is along the $x$ direction and the zigzag boundary is along the $y$ direction. Red and cyan rectangles mark the supercells along armchair and zigzag boundaries, respectively. (b) Input intensity pattern of the armchair edge state-I and armchair edge state-II. (c1, d1, e1) Output beams of the corresponding armchair edge state-I, armchair edge state-II and beams with a uniform phase distribution after $20\ mm$ propagation, whose phase distributions and Fourier spectra are shown in insets. The solid and dashed lines mark the center and edge of the 1D BZ in the insets of the Fourier spectra. The white shadows in (e1) mark the sites that initially have zero-amplitude. (c2, d2, e2) Corresponding simulation results of the armchair edge state-I, armchair edge state-II, and beams with a uniform phase distribution after long-distance propagation ($Z = 150\ mm$).

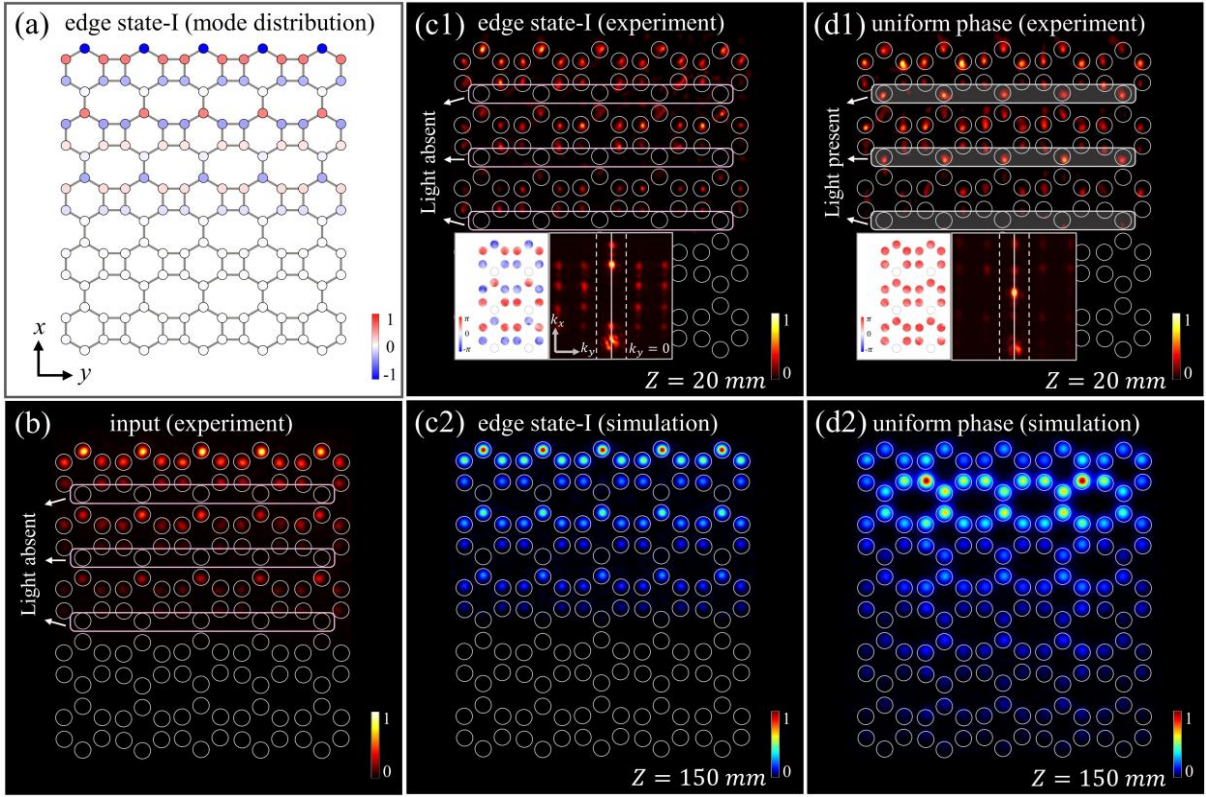

**Figure 4. Experimental demonstration of zigzag edge state-I in the photonic BPN.** (a) Mode distribution of zigzag edge state-I. (b) Input intensity pattern. (c1, d1) Output beams after $20\ mm$ propagation of zigzag edge state-I (c1) and uniform phase beams (d1). Insets show the phase distributions and Fourier spectra, with solid and dashed lines marking the center and edge of the 1D BZ. White shadows in (d1) highlight the initially zero-amplitude sites. (c2, d2) Corresponding simulation results for long-distance propagation ($Z = 150\ mm$).